# Modified self-consistent harmonic approximation and its application to two-dimensional easy-axis quantum ferromagnets

## D. V. Spirin[*]

*V.I. Vernadskii Taurida National University, Yaltinskaya st. 4, Simferopol, 95007, Crimea, Ukraine*

**PACS**: 75.10.Hk; 75.40.Cx; 75.10.Jm.

**Abstract.** In the paper we describe the modification of self-consistent harmonic approximation for quantum $S=1$ systems. This method has a number of advantages in comparison with usual SCHA. We apply the method to two-dimensional ferromagnets with easy-axis exchange or single-site anisotropy. The results are in good agreement with Monte-Carlo simulations and pure-quantum SCHA.

## I. INTRODUCTION

Low-dimensional systems attract much interest recently. Theoretically ultrathin magnetic films are usually described as two-dimensional (2d) systems. It is known that 2d isotropic ferromagnets cannot sustain long-range magnetic ordering. The account of easy-axis anisotropy leads to essential change of the magnetic properties: the long-range order is stabilized and easy-axis 2d ferromagnets may undergo a phase transition of the Ising type with finite and nonzero transition temperature [1].

Influence of easy-axis (exchange or single-site) anisotropy in 3d ferromagnets has been well studied [2]. As to 2d systems, the theoretical investigations were carried out mainly during last decade. Classical 2d easy-axis model was investigated by Monte-Carlo method [3], projected dynamics method [4], self-consistent harmonic approximation (SCHA) [5]. For description of the properties of 2d quantum ferromagnets the Green's function formalism [6,7], quantum SCHA [8], and pure-quantum SCHA (PQSCHA) [9] were used.

---

[*] Corresponding author. Home address: Kievskaya st. 121A, apt.69, Simferopol, 95034, Crimea, Ukraine. E-mail: spirin@crimea.edu, spirin@tnu.crimea.ua

In this paper we present the modification of SCHA for quantum systems proposed by authors [8]. Let us qualitatively describe the main idea of their method. SCHA consists in resorting of Hamiltonian through bose operators using, for example, Dyson-Maleev transformation. Taking into account four-particle interactions, one may calculate their contribution into two-particle terms (for details, see Ref. [8] and Refs. therein). The couplings of the model are replaced by quadratic interactions whose strength is then optimized. As a result one obtains self-consistent equations; solving them one can find the magnetization, Curie temperature, etc. Notice, that the method described in Ref. [8] is valid only for ordered magnets (i.e. easy-axis model, for instance), while the classical SCHA [5,10] can be applied to ordered systems, disordered ones (classical 2d isotropic Heisenberg ferromagnet) or those with topological order (2d XY or planar rotator models). The results obtained in Ref. [8] for 2d quantum ferromagnet with easy-axis exchange anisotropy are in good agreement with those obtained using Green's function technique [6].

The advantage of this method is its simplicity. However, we would like to emphasize that SCHA has also some drawbacks. First one is that the Hamiltonian obtained by implying the quantization procedure (Dyson-Maleev transformation) is non-Hermitian. Actually, the four-particle term appearing after this transformation has the form $a_1^+ a_2^+ a_2 a_2$, and no terms like $a_1^+ a_1^+ a_1 a_2$ emerge (*1,2* denote the lattice site numbers). Therefore this Hamiltonian describes some non-physical quantity. One may hope that the error caused by this is not so important. This is a usual lack, the same one can find in modified spin-wave theory, for example [11].

The second important point is that Dyson-Maleev transformation enables to account only low-energy excitations (*a*-magnons). Generally speaking, this is valid only for $S = 1/2$. For spin-*S* ferromagnet there are *2S+1* energy levels and *2S* different magnetic excitations may exist. Account of only low-frequency magnon branch in quantum case ($S \geq 1$) may lead to serious error, however in some cases the contribution from others can be discarded with great precision. One example of the case when all magnon branches must be taken into account is considered in Section III.

We would like also to criticize the interpretation of the results given in Ref. [8]. For spin-one Heisenberg ferromagnet with easy-axis exchange anisotropy Leonel and Pires found that dependence of Curie temperature on anisotropy parameter was linear. It was supposed that in the limit of large anisotropies one obtains a continuos spin Ising model. The critical temperature was estimated by calculating the slope of the $T_C(K)$ line, where $K$ is the anisotropy parameter. In Ref. [8] it was found that $T_{C,I}=2J$, and this temperature was compared with the result for 2d Ising model $T_{C,I}=2.27J$ calculated by Onsager.

We should empathize that Leonel and Pires [8] considered *spin-one quantum* ferromagnet, for which the projection of spin on *0Z* axis may be equal to *1, 0* or *–1*. The comparison of their result with that obtained by Onsager for Ising model (the projection of spin may be equal to *0.5, –0.5,* or *–1, 1* etc – it is no matter because one should renormalize the exchange strength respectively) seems to be very strange. The agreement of these temperatures would be more natural if the authors had compared Onsager result with theirs for *S=1/2*. However, simple numerical calculation (see Ref. [8], Eqs.(7)-(9)) shows that for spin one-half 2d ferromagnet in the limit of large anisotropy the slope (and, as consequence, the Curie temperature) is near to *1*. Moreover, for *quantum spin-one* Ising ferromagnet recent high statistics Monte-Carlo study gave $T_{C,I}=1.69\ J$ [12], in disagreement with the results of Refs. [6] and [8]. Although in Ref. [6] Curie temperature $T_{C,I}=2\ J$ for *S=1* 2d ferromagnet (in the Ising limit) was also obtained, we should mention that these authors used the *simplest* decoupling procedure (random-phase approximation). It is not clear this method is valid for 2d magnets. One may conclude that both SCHA [8] and RPA [6] fail to describe the Ising limit $K/J \to \infty$. Therefore we do not know whether these methods are valid in the whole anisotropy range.

One needs the theory which would be free of all drawbacks mentioned above. We propose such method that will be referred to as modified SCHA. The detailed calculations and explanation will follow in the next Section.

## II. 2D FERROMAGNETS WITH EXCHANGE ANISOTROPY

For quantum systems it is convenient to use the Hubbard operators method and bosonization technique (details can be found in Refs. [13,14] and Refs. therein). As a result of bosonization, one obtains the Hamiltonian resorted through *2S* types of bose operators. The resulting Hamiltonian is non-Hermitian, of course. Besides, the spin operators (and Hubbard ones as well) act in the *2S+1* – dimensional space. Using bosonization [14] one rewrites Hamiltonian through bose operators acting in infinite-dimensional space. Therefore there appear non-physical states such as $|\pm(2S+1+i)\rangle$, $i > 0$.

To cut off these states in Ref. [14] it was proposed to use the metric operator. Acting by this operator on Hamiltonian we automatically obtain a new one, which is already Hermitian. If one takes into consideration only two-particle terms then there is no need in this operator: it gives corrections into four-particle interactions (and interactions of higher order) only.

Let us shortly describe this technique. Consider 2d Heisenberg ferromagnet with easy-axis exchange anisotropy:

$$H = -J\sum_{1,2}\mathbf{S}_1\mathbf{S}_2 - K\sum_{1,2}S_1^z S_2^z. \qquad (1)$$

Here *J, K>0*, the summation is taken over nearest neighbors of site *1* (each bond is taken once) on a simple square lattice with coordination number *z=4*. *K* measures the strength of anisotropy and in the limit *K=0* we have an isotropic system with no long-range magnetic order [1].

Following Refs. [13], we separate out the mean field from Eq. (1). Thus we obtain the single-site Hamiltonian:

$$H_0 = -z\langle S^z \rangle (J+K)S^z. \qquad (2)$$

For *S=1* we take the low-temperature value: $\langle S^z \rangle = 1$. It is easy to solve the Schrödinger equation with Eq. (2), and find the energy levels and eigenfunctions.

Because the magnetic ion has *three* energy levels (we denote them as *1, 0, -1*), one obtains such relation of Hubbard operators with spin ones [13]:

$$S^+ = \sqrt{2}\left(X^{10} + X^{0-1}\right), \quad S^z = X^{11} - X^{-1-1}, \qquad (3)$$

By substituting Eq. (3) into Eq. (1) one can obtain the Hamiltonian rewritten through Hubbard operators, and the single-site part (2) becomes diagonal (one should keep in mind that interactions of Ising type like $S^z S^z$ in Eq. (1) are replaced by Eq.(2)).

Let us apply the bosonization procedure* [14]. We resort Hubbard operators through bose ones:

$$X_1^{10} = (1 - a_1^+ a_1 - b_1^+ b_1)a_1, \quad X_1^{01} = a_1^+, \quad X_1^{1-1} = (1 - a_1^+ a_1 - b_1^+ b_1)b_1,$$

$$X_1^{-11} = b_1^+, \quad X_1^{0-1} = a_1^+ b_1, \quad X_1^{-10} = b_1^+ a_1, \quad X_1^{00} = a_1^+ a_1, \qquad (4)$$

$$X_1^{-1-1} = b_1^+ b_1, \quad X_1^{11} = 1 - a_1^+ a_1 - b_1^+ b_1.$$

There emerge two types of magnons (*a* and *b*-magnons), which correspond to transitions $|1\rangle \leftrightarrow |0\rangle$ and $|1\rangle \leftrightarrow |-1\rangle$, respectively.

In Refs. [14] there was introduced the metric operator:

$$\overset{\otimes}{F} = 1 - \frac{1}{2}\sum_f a_f^+ a_f^+ a_f a_f - \frac{1}{2}\sum_f b_f^+ b_f^+ b_f b_f - \sum_f a_f^+ b_f^+ b_f a_f - \ldots \qquad (5)$$

These authors built it so that after action of this operator on any other Hubbard operator or their combination one obtains that i) $X^{pq} = \left(X^{qp}\right)^+$ holds for all *p, q* (from Eqs. (4) one can see this condition does not hold) and ii) it cuts off all nonphysical states.

For example, for spin operators one has:

$$\overset{\otimes}{F} S^- = \sqrt{2}\left(a - a^+ aa - b^+ ba + a^+ b + \ldots\right),$$

$$\overset{\otimes}{F} S^+ = \sqrt{2}\left(a^+ - a^+ a^+ a - b^+ ba^+ + ab^+ + \ldots\right), \qquad (6)$$

$$\overset{\otimes}{F} S^z = 1 - a^+ a - 2b^+ b + \ldots$$

However, this method has a serious lack. Namely, the commutation relation for spin operators $[S^-, S^+] = 2S^z$ does not hold even in the first order in $n$, where $n$ is the number of magnons of $a$ or $b$-type at the lattice site. One can easily check out this statement by calculating $\left[\overset{\otimes}{F} S^-, \overset{\otimes}{F} S^+\right]$, where $S^i$ are the components of spin operator rewritten through bose ones using Eqs. (3), (4). Namely, in $n^1$ order:

$$\left[\overset{\otimes}{F} S^-, \overset{\otimes}{F} S^+\right] = 2(1 - 3b^+ b - 3a^+ a) \neq (2S^z = 2(1 - 2b^+ b - a^+ a)). \tag{7}$$

---

* The transformation like Eq. (4) was proposed in F. P. Onufrieva, Sov. Phys. – JETP, **89**, 2270 (1985). However, for the first time the metric operator was derived in Refs. [14].

The reason is that the authors [14] had the aim to satisfy i) and ii) conditions *only*. However, ii) condition is not so important for *calculations*, because it leads to that the metric operator Eq. (5) has *infinite* number of terms. When one solves some problem in the four-particle approximation, for instance, one should keep only the four-particle terms in Eq. (5). Therefore the non-physical states will appear *in any way*. The price one pays building metric operator that would cut off all non-physical states is that the commutation relation for spin operators does not hold.

We build the bosonization procedure in different way. Let us require: i') $S^+ = (S^-)^+$, $S^z = (S^z)^+$ and ii') all commutation relations for spin operators must hold *up to the first order in $n$*. One may find the only one representation:

$$S^- = \sqrt{2}\left(a - \frac{1}{2} a^+ aa - \frac{1}{2} b^+ ba + a^+ b\right),$$

$$S^+ = \sqrt{2}\left(a^+ - \frac{1}{2} a^+ a^+ a - \frac{1}{2} b^+ ba^+ + ab^+\right), \tag{8}$$

$$S^z = 1 - a^+ a - 2b^+ b.$$

The advantage of this transformation which is valid in $n^1$ order is that the Hamiltonian rewritten through Eqs. (8) becomes Hermitian. The next step is to substitute Eqs. (8) into Eq. (1) and keep only four-particle terms.

The Hermitian Hamiltonian having been obtained, we use usual SCHA [8]. We find the effective two-particle energy:

$$H^{(2)} = \sum_k \varepsilon_{a,k} a_k^+ a_k + \sum_k \varepsilon_{b,k} b_k^+ b_k, \tag{9}$$

$$\varepsilon_{a,k} = z \begin{pmatrix} J + K - J\gamma_k + (2J\gamma_k - J - K)f_0 + (2J - \gamma_k J - \gamma_k K)f_\delta \\ + (J\gamma_k - 2J - 2K)g_0 - J\gamma_k g_\delta \end{pmatrix},$$

$$\varepsilon_{b,k} = z(2J + 2K - 2(J + K)f_0 + J(1 - \gamma_k)f_\delta - 4(J + K)g_0 - 4\gamma_k(J + K)g_\delta),$$

where $f_0, f_\delta, g_0, g_\delta$ are determined as:

$$f_\delta \equiv \frac{1}{N}\langle a_1^+ a_2 \rangle = \frac{1}{(2\pi)^2} \int_{-\pi}^{\pi} \int_{-\pi}^{\pi} \frac{\gamma_k}{\exp(\varepsilon_{a,k}/T) - 1} dk_x dk_y, \quad f_0 = f_\delta|_{\delta=0}, \quad (10)$$

$$g_\delta = \frac{1}{N}\langle b_1^+ b_2 \rangle, \quad g_0 = g_\delta|_{\delta=0}$$

and $\gamma_k = \frac{1}{2}(\cos k_x + \cos k_y)$ is the structure factor (we take $\delta_{x,y} = 1$. For $\delta \to 0$ in Eqs. (10) we have $\gamma_k \to 1$). The spectra corresponding to ideal gas of magnons (two-particle terms in Eq. (9)) coincide with those of Refs. [6,8].

One can also calculate dependence of magnetization on temperature by averaging Eq. (8):

$$m = 1 - f_0 - 2g_0. \quad (11)$$

The equations (9), (10) should be solved self-consistently by iteration procedure. We present the result of numerical calculation at Fig. 1, where the dependence of Curie temperature on $K$ at large anisotropies ($K \geq 4J$) is shown (solid line). This temperature is found as the point at which equations (9), (10) have no physical solution. The dashed line corresponds to the results of Leonel and Pires [8], dotted line was obtained by authors [9] in the frameworks of PQSCHA (Calculations for easy-axis 2d ferromagnet are presented at Fig. 9 of Ref. [9]. To compare their results with ours one should remember that they define material parameters and reduced temperature in other way).

First consider the case of Ising limit. The dependence of $T_C$ on $K$ is linear. For $K/J \gg 1$ we recover a spin-one Ising Hamiltonian with $T_{C,I}=1.69\ J$ (high statistics Monte-Carlo simulations) or $T_{C,I}=1.72\ J$ (Monte-Carlo single-spin-flip technique) [12]. One can estimate this temperature calculating the slope of the solid line $T_C(K)$. We obtain $T_{C,I}=1.67\ J$. This is in good agreement with the results of Monte-Carlo simulations [12] and in contradiction with the results of authors [6] and [8]. Our result is also in agreement with PQSCHA Ref. [9] where the estimation $T_{C,I}=1.73\ J$ was obtained (Fig. 6. of Ref. [9]).

For finite *K* our phase diagram is much closer to that obtained with PQSCHA (dotted line) than dashed line (SCHA, Ref. [8]). Because both our modified SCHA and PQSCHA give Curie temperature in the Ising limit very near to computer simulations, the question remains open which method is better.

Finally, we should note that in the case considered in this Section the effect of the high-frequency magnon branch is important. If one discards with *b*-magnons, we obtain the slope close to *1.8 – 2* and Curie temperatures become somewhat higher than those shown at Fig.1.

An example of the system when the account all magnon branches leads to interesting effect is presented in the next Section.

## III.  2D FERROMAGNETS WITH EASY-AXIS SINGLE-ION ANISOTROPY

Consider influence of single-ion anisotropy. The exchange Hamiltonian we choose in isotropic form and it coincides with Eq. (1) setting *K=0*. The single-ion term looks like:

$$H_a = -\frac{\beta}{2}\sum_{l}\left(S_l^z\right)^2, \tag{12}$$

with $\beta > 0$. Surely, *S>1/2;* we consider the case *S=1*. Let us apply the method of bosonization described above.

The results are quite similar to Eqs. (9), however, instead of *K* we should write $\beta/2$. We also mention that while rewriting Eq. (12) through bose operators (Eq. (8)), one must keep *only two-particle terms*. The reason is that our approximation is valid only in the first order in *n for the lattice site.* Thus, we hold terms like $a_1^+ a_1^+ a_1 a_2$ emerging from exchange part, while $a_1^+ a_1 a_1^+ a_1$, appearing from (12) give contribution into $a_1^+ a_1$ only (normal ordering procedure is used, of course).

Derivation of the result is not difficult, therefore we do not reproduce it here.

This model is of interest, because the energy levels of a magnetic ion have the form (see Refs. [13]): $E_1 = -\frac{\beta}{2} - zJ$, $E_0 = 0$, $E_{-1} = -\frac{\beta}{2} + zJ$. For sufficiently large anisotropies $\beta > 2zJ$ the inversion of energy levels takes place: $E_{-1} \to E_0$, $E_0 \to E_{-1}$. In fact, this means that energy of *b*-magnons becomes lower than that of *a*-magnons. Generally speaking, Dyson-Maleev representation becomes inadequate already for $\beta \propto zJ$. In such a case one must not discard with *b*-excitations anymore because they have the energy comparable with the energy of low-frequency magnon branch. This effect is displayed at Fig. 2, where we present numerical results as $T$-$\beta$ phase diagram.

Critical temperature depends on anisotropy linearly, however, at the point $\beta_c = 2zJ$ the slope slightly decreases. For $\beta < 2zJ$ we have $\frac{\Delta T_C}{\Delta \beta} \approx 0.033$, while for $\beta > 2zJ$ the we have $\frac{\Delta T_C}{\Delta \beta} \approx 0.016$ that is, two times lower. This is related with the fact that the magnetization for *S=1* depends on $f_0$ and $g_0$ (these functions describe the contribution from *a*- and *b*-magnons into spin deviation, respectively) in different ways (see Eq. (11)). Below $\beta_c$ the *a*-magnons have lower energy. However, for higher values of anisotropy their energy becomes higher, and *b*-magnons give larger contribution into spin deviation. Because *f* and *g* functions have different multipliers (*b*-magnons give contribution two times higher) in Eq. (11), the slope of the line is twice lower.

## 4. CONCLUSION

The modified SCHA we present here has some advantages in comparison with usual SCHA [8]. Let us summarize the results. Because the Hamiltonian which we start with to use SCHA is Hermitian, we believe that the obtained results should describe the properties of a system under study much better. Surely, any modification of SCHA (and ours as well) cannot be free of some drawbacks usual for this approximation [5]. It does not take into account strong coupling effects at high temperatures and neglects the kinematical interaction. Generally speaking, there is no theory which could predict the value of Curie temperature with great precision taking into account high-temperature fluctuations in two dimensions, except the renormalization group theory, of course. However, we think, that the modified SCHA we propose gives good estimations.

Our results for *S=1* 2d ferromagnet with exchange anisotropy are in contradiction with the results of Refs. [6] and [8]. However, for large anisotropies we obtain Curie temperature which is in good agreement with Monte-Carlo studies [12] and PQSCHA [9].

The transformation (7) takes into account *a* and *b*- excitations for spin-one ferromagnet. In the case of large single-ion anisotropy the account of high-frequency magnons is important and leads to the effect shown at Fig. 2.

One more advantage is that we obtain Hermitian Hamiltonian, Eqs. (8) having been used. Dyson-Maleev representation leads to non-Hermitian Hamiltonian. This may be essential if one applies the transformation to antiferromagnets (the coefficients at *aa* and $a^+a^+$ after SCHA will be different), easy-plane ferromagnets, etc.

We would like to mention that for spin one-half ferromagnet our transformation (8) will be almost the same as Dyson-Maleev one. Implying (8) (without *b*-magnons, all corresponding terms should be dropped) we obtain Hermitian Hamiltonian, but SCHA gives the same results as in Ref. [8].

In conclusion, our approach (modified SCHA) is likely to be more relevant for description of properties of quantum magnetic systems with $S >1/2$, than SCHA.

**ACKNOWLEGEMENTS**

Author thanks Ministry of Ukraine and Ministry of Education and Science of Ukraine (Grant 235/03) for the financial support.

Figure captions.

Fig. 1. Phase diagram *temperature-anisotropy* for two-dimensional spin-one ferromagnet with exchange anisotropy. Dashed line: results of Ref. [8], dotted line: Ref. [9].

Fig. 2. Phase diagram *temperature-anisotropy* for two-dimensional spin-one ferromagnet with single-site anisotropy.

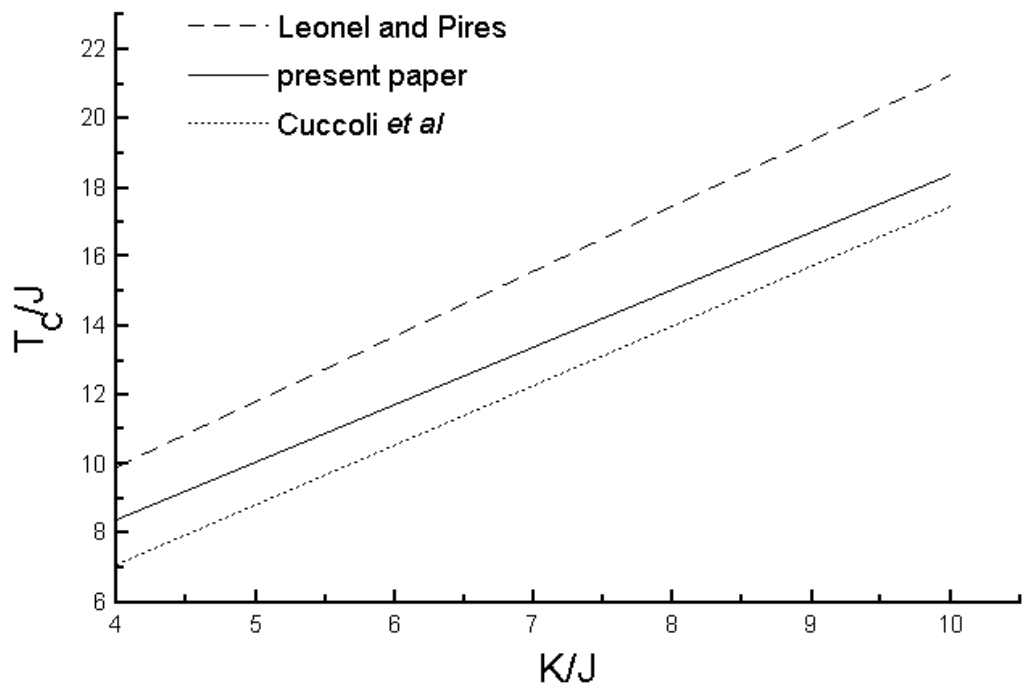

Fig.1.

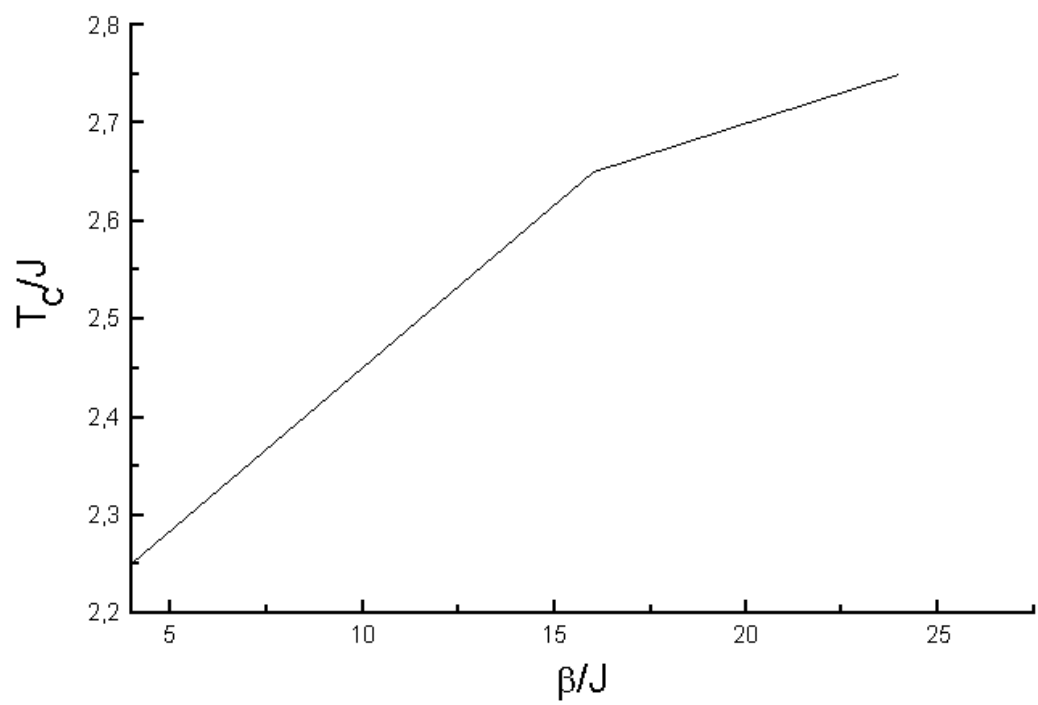

Fig.2.